\input{aipcheck}

\documentclass[
    ,final            
  ]
  {aipproc}

\layoutstyle{6x9}

\begin{document}

\title{Neutrino flavour conversion and supernovae}


\author{Cristina Volpe}{
  address={Institut de Physique Nucl\'{e}aire Orsay, CNRS/IN2P3, F-91406 Orsay
cedex, France}
}

\begin{abstract}
We summarize the recent developments in our understanding of neutrino flavour conversion in core-collapse supernovae and discuss open questions.
\end{abstract}

\maketitle

\noindent
The existence of neutrino mixings  and non-zero neutrino masses is responsible for the neutrino oscillation phenomenon -- these elusive particles change their flavour when they travel, as first discovered by Super-Kamiokande \cite{Fukuda:1998mi}. Theoretically,
accounting for neutrino oscillations requires the introduction of a unitary matrix relating the neutrino flavour to the mass basis, the Maki-Sakata-Nakagawa-Pontecorvo (MNSP) matrix, analogue to the Cabibbo-Kobayashi-Maskawa matrix in the quark sector. If  three neutrino families are assumed, the MNSP matrix depends upon
three mixing angles and three (two Majorana and one Dirac) phases. The former are now measured with good precision, the best being the third neutrino mixing angle determined by the T2K \cite{Abe:2011sj}, Double-Chooz \cite{Abe:2011fz}, Daya-Bay \cite{An:2012eh} and RENO \cite{Ahn:2012nd} experiments. An important open question is the value of the phases, since non-zero Dirac or Majorana phases render the MNSP matrix complex and introduce CP violation in the lepton sector. In particular, the T2K and NO$\nu$A experiments
are capable of covering a (small) fraction of the Dirac phase values that might indeed include the true value. If this is not the case, the investigation of
a large fraction of the Dirac phase values might ask for the construction of large size detectors such as Hyper-K, or experiments like Daedalus \cite{Conrad:2009mh}. Long term facilities, that have been discussed, include beta-beams \cite{Volpe:2006in} and neutrino factories. 

Neutrino oscillations in vacuum is an interference phenomenon among the mass (or propagation) eigenstates that acquire different phases, while they travel 
from the neutrino source to a detector. However, the presence of matter modifies
the way neutrinos change their flavour in a significant way.
Note that the neutrino interaction with the matter composing a medium is usually accounted for, at the mean-field level, by implementing an effective term to the neutrino Hamiltonian. Such contribution is linear in the weak coupling constant and proportional to the matter number density.
Its inclusion produces a resonant flavour conversion phenomenon known as the Mikheev-Smirnov-Wolfenstein (MSW) effect \cite{Wolfenstein:1977ue,Mikheev:1986gs}. The occurrence of the resonance and the efficiency of the flavour conversion depend upon the adiabaticity of the evolution at the resonance,which in turn is determined by the specific matter density profile of the medium and the neutrino properties (energies, mixing angles, mass-squared differences and their sign). In particular, it is now established that the MSW effect is at the origin of the "solar neutrino deficit problem" (see e.g. \cite{Collaboration:2011nga}) first observed by R. Davis' pioneering experiment \cite{Davis:1964hf}. In particular,
the occurrence of an MSW resonance for electron neutrinos in the Sun tells us one of the mass-squared differences' sign ($\Delta m^2_{12} > 0$).

At present, the ensemble of the neutrino oscillation experimental results is well described within the three neutrino oscillation framework. However, some anomalies have been identified, in particular, in the reactor data (the "reactor" anomaly) \cite{Mention:2011rk} and in a measurement with a static source (the "Gallium" anomaly) \cite{Giunti:2010zu}. The results of the accelerator-based MiniBOONE experiment also present puzzling aspects that still need to be clarified \cite{AguilarArevalo:2012va}. If confirmed, the interpretation of these anomalies will require going beyond the employed theoretical framework and implementing, for example, 
the inclusion of a fourth sterile neutrino, or of some other exotic possibility.
Numerous projects have been proposed to address the hypothesis of
 a fourth sterile neutrino having a large squared-mass difference, in the eV$^2$ range. These comprise experiments using either a detector located at short baselines (e.g. \cite{Anderson:2012pn,Dwyer:2011xs}), or a static source inside already existing neutrino detectors such as Borexino or Kamland  as in \cite{Cribier:2011fv}, or $\beta$-decaying nuclei produced at radioactive ion beam facilities and impinged on a target inside a $4 \pi$ detector  \cite{Espinoza:2012jr}. Clarifying 
the origin of these anomalies constitutes one of key open issues. 
Other crucial open questions include the determination of the neutrino mass scale, of the neutrino (Dirac or Majorana) nature and of the neutrino mass hierarchy. 
The KATRIN experiment \cite{Bonn:2010zz} is about to explore the sub-eV range for the neutrino effective mass; while double-beta decay measurements are also sensitive to the nature. 

Understanding how neutrinos modify their flavour in media is a fascinating theoretical problem, also crucial for observations that range from solar and core-collapse supernova neutrino experiments, to both stellar and primordial nucleosynthesis abundances. A variety of mechanisms can be engendered, depending upon the specific medium neutrinos are traversing - a smooth matter density profile as the one in our Sun, the abrupt matter change present in the Earth, the turbulent media of violent events as core-collapse supernovae, the accretion-disks around black-holes (AD-BH), and expanding environments as the Early Universe. Here we will focus upon the core-collapse supernova case and describe flavour conversion effects associated with the fact that neutrinos meet lots of neutrinos when traversing these massive stars.

Core-collapse supernovae occur at the end of the life of massive stars. 
While our Sun continuously emits electron neutrinos, core-collapse supernovae produce neutrinos and anti-neutrinos of all flavours in a burst lasting about 10 seconds. Such explosions produce an enormous amount of neutrinos since $99 \%$ of their gravitational energy is taken by about 10$^{57}$ neutrinos with 10 MeV average energy. These expectations have been roughly confirmed by the electron anti-neutrino events, measured during the explosion of the SN1987A, located in the Large Magellanic Cloud at 50 kpc (see e.g. \cite{Suzuki:2008zzf,Pagliaroli:2008ur} and references therein). Theoretically, a significant step forward has been done, with the increase in complexity of core-collapse supernova simulations,
that are nowadays based upon multidimensional computations and include convection, realistic neutrino transport and dynamical instabilities - the Standing Accretion Shock Instability (see e.g. \cite{Bruenn:2010af}). Although a consensus is not yet reached on the relative importance of these different aspects and on the precise mechanism, different groups are now capable of obtaining succesfull explosions for various progenitor masses. This progress makes us hope that an answer to the longstanding problem of understanding how these massive stars explode, might find an answer in the coming decade.

 
Clearly, key information on the explosion mechanism and on neutrino properties could be gained if novel experimental data are obtained. Core-collapse supernovae in our Galaxy are rare events, they become as frequent as abount one event per year at distances as large as 1-4 Mpc \cite{Ando:2005ka} (obviously only a bunch of events could be measured from far events). The network of running neutrino detectors around the world can measure a supernova explosion if it occurs in our galaxy. If a supernova explodes tomorrow,
running detectors can collect several thousands events altogether; while if one of
the large (megaton) size detectors under study \cite{Autiero:2007zj}  is built, up to $10^5$ events can be collected.
So many events would give us the picture of the explosion, seen
with neutrinos : their emission closely follows the different phases under which a massive star dies. In order to extract as much as information as possible on neutrino properties and the supernova dynamics, one needs a network of
detectors based upon various technologies, and having different
thresholds. Ref.\cite{Vaananen:2011bf} shows examples of how the presence of one- and two-neutron emission thresholds in a lead-based supernova detector like HALO  can significantly reduce the number of degeneracies. 
Supernova neutrino observatories exploiting nuclei to detect neutrinos, such as HALO, would benefit from the realization of low energy neutrino scattering measurements of neutrino-nucleus cross sections at spallation sources \cite{Avignone:2003ep,Lazauskas:2010rh}, or at a low energy beta-beam \cite{Volpe:2003fi}.  

The diffuse supernova neutrino background, produced by explosions at different redshifts, encodes information on the star formation rate as well. 
Studies are ongoing to determine if an upgraded neutrino detector technology, based upon the addition of Gadolinium to a water Cherenkow detector, allows to reach the required background suppression (see \cite{Beacom:2010kk} for a review). Predictions indicate that large size detectors should have the sensitivity for such a discovery. For example, in ten years one expects on average about 340 events in 440 kton water Cherenkov detector, about 60 events in 100 kton liquid argon and about 100 events in 50 kton liquid scintillator \cite{Galais:2009wi}. 
Finally, neutrinos play a role in stellar nucleosynthesis processes (neutrino nucleosynthesis, $r-$process, $\nu p$ process). Although such processes strongly depend upon the specific astrophysical conditions and site, 
as well as nuclear physics inputs, they are often sensitive to neutrino properties,
such as the presence of a fourth sterile neutrino \cite{McLaughlin:1999pd}.
In conclusion,
accurate predictions of the supernova neutrino signal, of the diffuse supernova neutrino background and of nucleosynthesis abundances require a precise understanding of how neutrino change their flavour in these explosive environments. This domain is being actively investigated since many novel
flavour conversion phenomena have been identified, while important open questions remain.  
 
\section{$\nu$ flavour conversion in core-collapse supernovae}

\begin{figure}
  \includegraphics[height=.2\textheight]{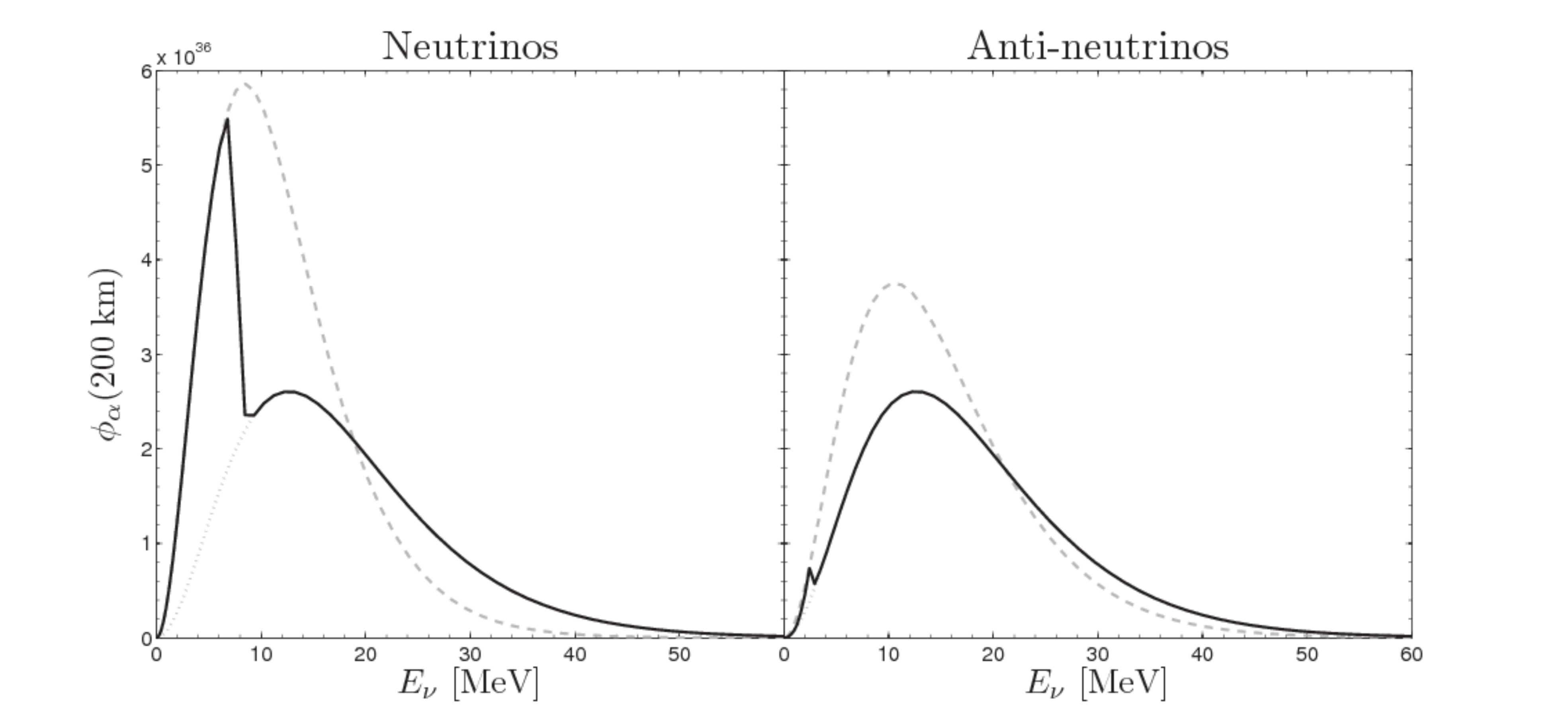}
  \caption{Electron neutrinos and anti-neutrino fluxes at 200 km from the neutrinosphere in a core-collapse supernova. The calculation includes a realistic one-dimensional supernova density profile and the neutrino interaction with neutrinos. The dotted (solid) lines present the fluxes in absence of (presence of) the $\nu-\nu$ interaction. The sharp flux change is characteristic of the spectral split phenomenon \cite{Galais:2011gh}.}
\end{figure}

\noindent
Neutrino flavor conversion in core-collapse supernovae has revealed to be complex. Three features produce characteristic flavour conversion phenomena : the presence of large neutrino number densities, of (front and reverse) shock waves \cite{Schirato:2002tg,Fogli:2003dw,Tomas:2004gr,Choubey:2006aq,Dasgupta:2005wn,Gava:2009pj,Galais:2009wi} and of turbulence \cite{Loreti:1995ae,Fogli:2006xy,Friedland:2006ta,Kneller:2010sc}.
Pantaleone first pointed out the
importance of the neutrino-neutrino interaction and the fact that it would introduce a non-linear refractive index \cite{Pantaleone:Gamma1292eq}. At the mean-field level the effective Hamiltonian includes a term which is proportional to the neutrino density matrix
(instead of the matter number density as in the MSW effect) to encode the fact
that neutrinos can change their flavour from the neutrinosphere to the point where they interact with another neutrino. It was recognized in  \cite{Samuel:1993uw} and recently in \cite{Sawyer:2005jk,Duan:2005cp,Duan:2006an} that the flavour evolution is drastically modified after the implementation of such a term. 
Three flavour conversion regimes might be present in nature. 
They consist first in a collective synchronization of the neutrinos with no flavor conversion, then in the occurrence of "bipolar" flavour oscillations and finally in a complete swap of the (anti) neutrino spectra above a critical energy  - the spectral split phenomenon (Figure 1) (for a review see \cite{Duan:2010bg}). 

A useful formalism to pin down the phenomena underlying the flavour conversion effects, observed in numerical calculations, is offered by the quantum mechanical analogy with effective spins. This consists in  associating to the neutrino amplitudes an effective spin $\bf{S}$, to the Hamiltonian an effective magnetic field $\bf{B}$, and solve a precession equation instead of the Schr\"odinger equation \cite{Cohen}. In particular, the third component of the spin (also named isospin) gives the neutrino survival probability, while the $x$- and $y$-components depend upon the neutrino mixings. This analogy was already used to describe vacuum oscillations as well as the MSW effect (see e.g. \cite{Kim:1987bv}).
(Note that, in these cases, the effective magnetic field only has real components.)
The MSW resonance occurs when the $z$-component of the magnetic field becomes
zero. An adiabatic evolution corresponds to the fact that the spin precession frequency is fast compared to the magnetic field evolution, so that the spin evolves with it through the resonance instead of lagging behind
(non-adiabatic solution).
The $\nu\nu$ interaction term 
has complex contributions because of its dependence upon the neutrino density matrix. This introduces a time-varying effective magnetic field in the plane.
Ref.\cite{Duan:2005cp} has shown that in the synchronization regime the total $\bf{S}$ precesses around $\bf{B}$ (almost aligned with the $z$-axis), dominated by the $\nu\nu$ interaction, so that no flavour conversion occurs. 
Using the analogy with a pendulum, in \cite{Hannestad:2006nj} it has been pointed out that, contrarily e.g. to the well known MSW effect, collective flavour conversion effects occur for any value of the third neutrino mixing angle, while such a parameter needs to be stricly non-zero. 
Bipolar oscillations can be associated to the evolution of a flavour pendulum \cite{Duan:2005cp,Duan:2007mv} and a gyroscopic pendulum \cite{Hannestad:2006nj} that makes both a precession and a nutation. 
This picture is further investigated in \cite{Wu:2011yi}.
Concerning the spectral split phenomenon associated with a sharp spectral swapping above a critical energy (Figure 1), in \cite{Duan:2007mv} 
the authors make the hypothesis that at the end of the synchronization and of 
the bipolar regimes, the neutrino evolution follows a collective precession mode until neutrino densities are low 
and at the late stage of this precession solution the stepwise swapping occurs.
The existence of a comoving frame is pointed out in the  early work \cite{Duan:2005cp}.  
In \cite{Raffelt:2007cb}, 
an explicit adiabatic solution is postulated in the comoving frame -- an MSW-like solution -- that is shown to behave like the spectral split solution. 
The split energy is explained by the (approximate) conservation of
the total neutrino lepton number (for the two-flavour case) \cite{Raffelt:2007cb}. The extension to three flavours, based upon
SU(3) is first studied in \cite{Dasgupta:2007ws}. Ref.\cite{Pehlivan:2011hp} shows that there exists an exact solution to the full two-neutrino problem (in the single-angle approximation and if the matter term is neglected). The authors put this neutrino Hamiltonian in relation with the 
class of Gaudin models and, in particular, the (reduced) BCS pairing Hamiltonian describing superconductivity. The spectral split phenomenon is understood in terms of the adiabatic evolution from the quasiparticle degrees of freedom in the high-density region where they coincide with the flavor eigenstates, to the vacuum, where they coincide with the mass eigenstates.

\begin{figure}
  \includegraphics[height=.2\textheight]{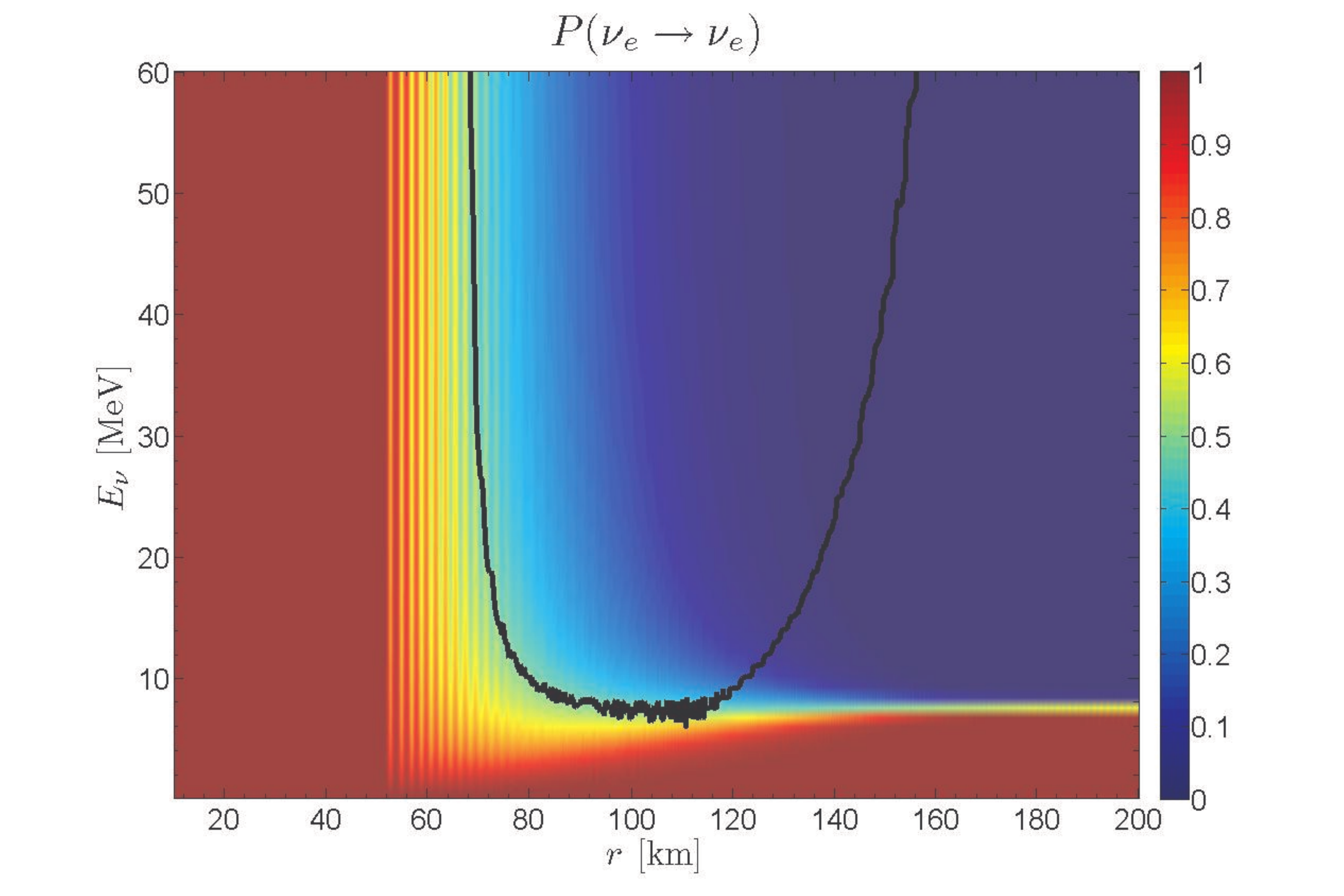}
  \caption{Three-dimensional contours of the electron neutrino survival probability as a function of neutrino energies (y-axis) and of distance in  a core-collapse supernova (x-axis) from the neutrinosphere set at 10 km. The black solid lines shows the neutrino energies and location in the supernova where the magnetic resonance conditions are fulfilled : $\Delta\omega = \omega-\omega_0 = 0$, $\Delta\omega / {\omega_1}\ll 1$. These coincide with the place and the neutrino energies that undergo the spectral split phenomenon  \cite{Galais:2011gh}. (Note that, when the solid line upturns, the conditions are not both fulfilled, so that no spin flip - or flavour change - occurs).}
\end{figure}

In a recent work, we have been showing that the spectral split phenomenon is a magnetic resonance (MR) phenomenon \cite{Galais:2011gh}. In its simplest realization,
the MR consists in a spin-flip occurring 
in presence of a constant and a time varying magnetic fields. Figure 2 shows the result of a two-flavour numerical calculation that fully includes the mixings, the matter, the neutrino-neutrino term and uses a realistic supernova density profile. The solid line corresponds to the neutrino energies and the location where the magnetic resonance conditions are fulfilled, 
namely $\Delta\omega = \omega-\omega_0 = 0$, $\Delta\omega / {\omega_1}\ll 1$, 
that coincide with the neutrino energies and the location in the supernova, 
where the spectral split phenomenon occurs. Note that, saying that a MR phenomenon occurs in particular implies the existence of an MSW-like solution in the comoving frame. This frame has been numerically identified (in the matter basis), as given by the average of the (fast varying) magnetic field. The results
of Ref.\cite{Galais:2011gh} give a consistent picture, in the flavour and in the matter basis, of the neutrino flavour evolution during the spectral split phenomenon. 

An important aspect, emerged with the $\nu\nu$ interaction inclusion in the neutrino flavour evolution simulations 
is the emergence of instabilities in flavour space.
Refs.\cite{Hannestad:2006nj,Duan:2007mv} have shown that bipolar oscillations start because of the presence of an instability. Indeed, such instabilities appear under various conditions, and can be present for any neutrino mass hierarchy, depending on the primary neutrino fluxes \cite{Dasgupta:2009mg} (multiple spectral splits), or in multi-angle calculations \cite{Sawyer:2008zs,Mirizzi:2012wp}. A linearized analysis of instabilities is proposed in \cite{Banerjee:2011fj}. Ref.\cite{Galais:2011jh} has shown that
the instability associated with the start of bipolar oscillations is due to 
a rapid growth of the matter phase.

A key question for observations is to determine how much of the flavour conversion effects described survive, when inputs (neutrino fluxes, neutrinospheres, matter density profiles) from multi-dimensional (exploding) supernova simulations are employed. For example, the presence of matter can introduce decoherence (see e.g. \cite{Duan:2007mv,EstebanPretel:2007ec}) and, eventually suppress these effects \cite{Chakraborty:2011gd}, during some phases of the explosion. 
One should be aware that flavour conversion phenomena induced by the $\nu\nu$ interaction occur in a region, deep in the star, while the MSW resonances, produced by the neutrino interaction with matter, occur in the low density layers.
This fact is particularly intriguing because on one hand such effects can be relevant for the explosion mechanism (depending upon their exact location
in realistic three-dimensional simulations), and for nucleosynthesis.
In fact, the significant impact on the electron fraction is pointed out in Ref.\cite{Balantekin:2004ug}  while Ref.\cite{Duan:2010af} has 
shown that predicting the $r$-process abundance require 
accurate descriptions of the $\nu\nu$ effects. 
For example Ref.\cite{Dasgupta:2011jf} has explored
a possible impact on the supernova explosion and found that, in principle, these
conversion effects occur outside the stalled shock enhancing heating by only a few percent.
However answering this key issue might require relaxing some of the approximations made in flavour conversion studies and realize demanding numerical calculations \cite{Cherry:2012zw}. 
 Much more work is definitely required to quantify the impact of flavour conversion on these open issues.

One motivation to study and observe core-collapse supernova neutrinos is the possibility to learn more about
key unknown neutrino properties. The neutrino mass hierarchy problem arises from the fact one of the sign is unknown : the third mass eigenstate can be the heaviest, or the lightest. The $\Delta m^2_{13}$ sign could be determined by exploiting
neutrino flavour conversion in matter in complementary ways, i.e. either in long-baseline accelerator experiments, or by measuring astrophysical neutrinos. One possibility is the observation of core-collapse supernova neutrinos produced in future explosions (see e.g. \cite{Gava:2009pj,Dasgupta:2008my}), or of athmospheric neutrinos going through the Earth and detected in ice or water Cherenkov detectors (see e.g. \cite{Bernabeu:2003yp,Akhmedov:2012ah}).
More precisely, it is shown in  \cite{Gava:2009pj} that, if the hierarchy is inverted, a dip or bump should be present in the positron time signal associated with a supernova explosion, depending upon the impinging neutrino energy, in 
a Cherenkov or scintillator detector. The same applies to the neutrino signal, if the hierarchy is normal. 

The value of the CP violating Dirac phase is one of the key unknowns that will be among the goals of the future neutrino experimental searches. If non-zero, it
can have an impact in astrophysical and cosmological environments (besides leptogenesis). Ref. \cite{Akhmedov:2002zj} first studied 
CP effects in the supernova context (with a negative result). 
The existence of possible CP violating effects in supernovae has been established in \cite{Balantekin:2007es} and \cite{Gava:2008rp} in presence of the neutrino-neutrino interaction. In particular, effects can arise due
to loop corrections or to physics beyond the Standard Model, or to any physics
breaking the factorization condition established in  \cite{Balantekin:2007es}.
A first quantification of the Dirac phase impact on the neutrino fluxes shows modifications at the level of $5-10 \%$. It is still too early to exclude that an enhancement of such effects could be present
and, maybe, have an impact on observations. Ref.\cite{Gava:2008rp} has studied for the first time
the possible effect of a non-zero Dirac phase on the neutrino degeneracy parameter and on the primordial light element abundances. 
By considering three active flavours only, modifications at the level of 1$\%$ of the primordial $^{4}$He abundance, due to a non-zero Dirac phase, are found.

In conclusion, the domain of supernova neutrinos is steadily progressing.
Exciting novel flavour conversion phenomena have been uncovered. Their understanding
 and interplay with realistic supernova modelling and nucleosynthetic simulations still require serious efforts. Mantaining a network of neutrino detectors, and possibly constructing a large-size one, might offer the unique opportunity to pin down the secrets of supernovae.


\bibliographystyle{aipproc}   

\end{document}